\documentclass{webofc}
\usepackage[varg]{txfonts}   % Web of Conferences font
\usepackage[utf8]{inputenc}
\usepackage{amsmath}
\usepackage{amsfonts}
\usepackage{amssymb}
\usepackage{graphicx}
\usepackage{xcolor}
\usepackage{dsfont}
\usepackage{cleveref}
\begin{document}
\title{A first sketch: Construction of model defect priors inspired by dynamic time warping}

\author{\firstname{Georg} \lastname{Schnabel}\inst{1}\fnsep\thanks{\email{georg.schnabel@physics.uu.se}} \and
        \firstname{Henrik} \lastname{Sjöstrand}\inst{1}
}

\institute{Division of Applied Nuclear Physics, Uppsala University}

\abstract{%
Model defects are known to cause biased nuclear data evaluations if they are not taken into account in the evaluation procedure.
We suggest a method to construct prior distributions for model defects for reaction models using neighboring isotopes of $^{56}$Fe as an example. A model defect is usually a function of energy and describes the difference between the model prediction and the truth. Of course, neither the truth nor the model defect are accessible. A Gaussian process (GP) enables to define a probability distribution on possible shapes of a model defect by referring to intuitively understandable concepts such as smoothness and the expected magnitude of the defect. Standard specifications of GPs impose a typical length-scale and amplitude valid for the whole energy range, which is often not justified, e.g., when the model covers both the resonance and statistical range. In this contribution, we show how a GP with energy-dependent length-scales and amplitudes can be constructed from available experimental data. The proposed construction is inspired by a technique called dynamic time warping used, e.g., for speech recognition.
We demonstrate the feasibility of the data-driven determination of model defects by inferring a model defect of the nuclear models code TALYS for (n,p) reactions of isotopes with charge number between 20 and 30.
The newly introduced GP parametrization besides its potential to improve evaluations for reactor relevant isotopes, such as $^{56}$Fe, may also help to better understand the performance of nuclear models in the future.
}
\maketitle
\section{Introduction}
\label{intro}
Evaluated nuclear data are important input for all kinds of nuclear physics applications.
It has been shown in the past, e.g., \cite{helgessonAssessmentNovelTechniques2017}, that evaluation techniques without proper account of potential model defects tend to underestimate uncertainties and bias results.
Acknowledging the problem and being in need of a solution, Bayesian procedures have been often pragmatically tuned to deal with the problem, e.g., by ad-hoc adjustments of the likelihood function or by rescaling the obtained posterior uncertainties to make them plausible.
A methodological elegant way is to introduce the information about model defects already as prior knowledge.
In the field of nuclear data evaluation, this approach was pioneered by Pigni and Leeb~\cite{pigni_uncertainty_2003}.
Since then, various improvements and suggestions have been made, e.g.,~\cite{leeb_consistent_2008} and it was recognized~\cite{schnabel_large_2015} that these developments essentially deal with the design of a covariance function for a Gaussian process (GP), e.g.,~\cite{rasmussen_gaussian_2006}.
GPs are flexible tools in the domain of non-parametric Bayesian statistics but commonly used specifications, such as the squared exponential covariance function, may not always be optimal for scenarios we encounter in nuclear data evaluation, e.g., quickly rising cross sections near thresholds or various degrees of predictive power of a nuclear model depending on the energy region where it is employed.
So far this problem has been addressed by tailoring the covariance function to the specific situation, e.g.,~\cite{schnabel_differential_2016,2018arXiv180300928S,helgessonFittingDefectNonlinear2017}, whose shape can be adjusted by a few so-called hyperparameters.

Our suggestion in this paper is to replace the typically very structured covariance function by a very flexible one and to infer its shape by taking into account reaction systems that can be considered similar, i.e., the same reaction channels of neighboring isotopes.
Formally, the amplitude and length-scale parameter---two numbers---of a standard squared exponential covariance function are replaced by an amplitude function and a metric function.
The introduction of the latter function was inspired by a technique called dynamic time warping used, e.g., in speech recognition~\cite{myersPerformanceTradeoffsDynamic1980} where the time axis is locally stretched and shrank to align different voice samples as good as possible.
In our case, it is exactly the same idea: To modify the effective distance between energies in order to align the Gaussian process as good as possible to the experimental data of all reaction systems at the same time.
We remark that a certain construction of a model defect based on the information of neighboring isotopes has already been suggested in~\cite{leeb_consistent_2008}.
A distinctive feature of our approach is the determination of the model defect by the maximization of a score function, i.e., the marginal likelihood.

This paper is structured as follows. In \cref{sec:dtwgp} we introduce the new parametrization of a GP which we call \textit{dynamic time warping GP} or DTW GP for short.
In \cref{sec:marlikemax} we derive the criterion to determine the amplitude function and metric function.
In \cref{sec:example} we demonstrate the feasibility of the data-driven approach to infer the model defect of TALYS for (n,p) reactions using data for $^{56}$Fe and neighboring isotopes.
Finally, in \cref{sec:summary} we summarize and conclude.
\section{DTW GP}
\label{sec:dtwgp}
The relation between experimental data and the model prediction can be written as
\begin{equation}
\label{eq:statmodel}
  \vec{\sigma}_\textrm{exp} = 
  \mathcal{M}(\vec{p}) +
  \vec{\varepsilon}_\textrm{def} +
  \vec{\varepsilon}_\textrm{exp}
\end{equation}
where $\vec{\sigma}_\textrm{exp}$ contains the experimental measurements, $\mathcal{M}(\vec{p})$ is the corresponding model prediction based on the set of model parameters $\vec{p}$, the vector $\vec{\varepsilon}_\textrm{def}$ contains the deficiency of the model, and $\vec{\varepsilon}_\textrm{exp}$ contains the errors of the measurements.
The unobservable truth $\vec{\sigma}_\textrm{true}$ is therefore given by both $\mathcal{M}(\vec{p}) + \vec{\varepsilon}_\textrm{def}$ and $\vec{\sigma}_\textrm{exp} - \vec{\varepsilon}_\textrm{exp}$.
The only accessible quantity in this model is $\vec{\sigma}_\textrm{exp}$.
The values in all other quantities are uncertain and we have to assign probability distributions to express our belief about the likelihood of possible realizations.
Once all prior probability distributions are defined, we can use Bayesian statistics to obtain estimates of all quantities involved.
In this section, we discuss the specification of a probability distribution for $\vec{\varepsilon}_\textrm{def}$.
The specifications of probability distributions for the other quantities follows in~\cref{sec:marlikemax}.

For the following discussion, we assume that $\vec{\sigma}_\textrm{exp}$ in \cref{eq:statmodel} contains angle-integrated cross sections $\sigma_{\textrm{exp},i}$ of a single reaction channel measured at incident energies $E_i$.
The corresponding model defect term $\vec{\varepsilon}_\textrm{def}$ can be specified as a Gaussian process, e.g.,~\cite{rasmussen_gaussian_2006}.
A Gaussian process is the generalization of a multivariate normal distribution to functions and is characterized by the marginal distributions over finite sets of function values:
Any finite set of function values at arbitrary locations is governed by a multivariate normal distribution.
A Gaussian process is therefore uniquely defined by a mean function $\mu(E)$ and a covariance function $k(E, E')$.
As the names imply, the former provides a mean value for any energy $E$ and the latter covariances between pairs of function values at arbitrary energies.
Expressed in terms of our setting: For any choice of incident energies $E_1, E_2, ...$ the probability distribution of associated cross sections follows a multivariate normal distribution.
We assume that the model prediction is a priori the most probable option, i.e., $\mu(E) = 0$.
A common choice for the covariance function is the so-called squared exponential form, e.g.,~\cite{rasmussen_gaussian_2006}, introduced in~\cite{schnabel_large_2015} for nuclear data evaluation,
\begin{equation}
\label{eq:sqrexpcovfun}
  k(E, E') = \delta^2 \exp\left( -\frac{1}{2} \frac{(E - E')^2}{\lambda^2} \right) \,.
\end{equation}
This form is a reasonable default but depending on the energy range and observable other forms may be more suitable.
Some alternatives have been employed and studied in~\cite{schnabel_differential_2016,2018arXiv180300928S,helgessonFittingDefectNonlinear2017}.
All of these alternatives, however, are still rather rigid in terms of their structure and therefore foremost good solutions for the observable and energy range for which they have been designed.
In this paper, we aim to construct a very flexible covariance function that is capable to adapt to any setting.
Therefore we suggest the covariance function
\begin{equation}
\label{eq:dtwcovfun}
  k(E, E') = \delta(E) \delta(E') \exp\left( -\frac{1}{2} \Big(m(E) - m(E')\Big)^2 \right) 
\end{equation}
where both the amplitude $\delta$ and length-scale $\lambda$ of the squared exponential form are replaced by an energy-dependent amplitude and metric function, respectively.
We call $m(.)$ a metric function because a distance (i.e. difference) between two energies gets transformed to another one.
We demand the triangle inequality to hold, which is a defining criterion for a metric.
Therefore, the function $m(.)$ must be monotonically increasing.
The idea we pursue in this contribution is the determination of the functions $\delta(.)$ and $m(.)$ in a data-driven way by taking into account the same reaction channel of neighboring isotopes.
The result for the (n,p) channel is displayed in~\cref{fig:optResults}.
The remainder of this and the next section deals with the parametrization of $\delta(.)$ and $m(.)$ and the method to infer its shape from the data.

We want a maximum of flexibility for the functions $\delta(.)$ and $m(.)$ and therefore suggest to define them as continuous piecewise linear functions.
A continuous piecewise linear function can be written as
\begin{equation}
\label{eq:pwlinearfun}
  f(E \,|\, \{z_i\}_{i=1..N}) =
  \sum_{i=1}^{N-1}
  \left(
  \frac{E_{i+1} - E}{E_{i+1}-E_i} \mathbf{z_i} +
  \frac{E - E_{i}}{E_{i+1}-E_i} \mathbf{z_{i+1}}
  \right) \,.
\end{equation}
The defining parameters of this function are the function values $\{z_i\}_{i=1..N}$ at energies $\{E_i\}_{i=1..N}$, i.e., $f(E_i) = z_i$.
With a dense enough energy grid, arbitrary functions can be well approximated.
The flexibility of functions parametrized as in \cref{eq:pwlinearfun} requires reasonable constraints on features of these functions, which can be regarded as prior knowledge.
Plausible constraints can be formulated in terms of minimal and maximal allowed changes per energy unit.
In order to properly formulate such constraints, we can make the variable transformation
\begin{equation}
  z_i(z_0, \Delta z_1, \cdots, \Delta z_N) = z_0 + \sum_{k=1}^{i} \Delta z_k \,.
\end{equation}
Using this variable transformation, we obtain
\begin{equation}
  g(E \,|\, z_0, \Delta z_1, \dots, \Delta z_N ) = 
  f(E \,|\, \{ z_i(z_0, \Delta z_1, \cdots, \Delta z_N) \}_{i=1..N} ) \,.
\end{equation}
With this definition, the piecewise linear functions parametrized in terms of differences of function values for $\delta(.)$ and $m(.)$ can be written as
\begin{equation}
\label{eq:transAmpAndMetric}
  \delta(E) = g(E \,|\, u_0, \cdots, u_N ) 
  \textrm{ and }
  m(E) = g(E \,|\, v_0, \dots, v_N ) \,.
\end{equation}
In the next section, we discuss how to determine the parameters $u_i$ and $v_i$ in a data-driven way by optimization.

\section{Marginal likelihood maximization}
\label{sec:marlikemax}
The determination the parameters $\{u_i\}_{i=1..N}$ and $\{v_i\}_{i=1..N}$ in \cref{eq:transAmpAndMetric} requires a score function to assess the performance of possible solutions.
In this section we derive one within the framework of Bayesian statistics.
First, we have to complete the statistical model in \cref{eq:statmodel} by specifying the missing probability distributions for the model parameters $\vec{p}$, and the experimental errors $\vec{\varepsilon}_\textrm{exp}$.
We assume multivariate normal distributions for both of them, i.e.,
\begin{equation}
  \vec{\varepsilon}_\textrm{exp} \sim
  \mathcal{N}(\vec{0}, K_\textrm{exp} )
  \textrm{ and }
  \vec{p} \sim \mathcal{N}(\vec{p}_0, K_\textrm{par} ) \,.
\end{equation}
For the model defect we have $\vec{\varepsilon}_\textrm{def} \sim \mathcal{N}(\vec{0}, K_\textrm{def})$ with the elements of the covariance matrix $K_\textrm{def}$ determined by \cref{eq:dtwcovfun}.
The assumption of a priori independence of $\vec{p}$, $\vec{\varepsilon}_\textrm{def}$, and $\vec{\varepsilon}_\textrm{exp}$ completes the specification of the statistical model.
In order to obtain analytic solutions, we further substitute the nuclear model by a first-order Taylor approximation 
\begin{equation}
\label{eq:linearmodel}
  \mathcal{M}_\textrm{lin}(\vec{p}) =
  \vec{\sigma}_\textrm{mod} + 
  S (\vec{p} - \vec{p}_0)
\end{equation}
with
\begin{equation}
    \vec{\sigma}_\textrm{mod} = \mathcal{M}(\vec{p}_0)
    \textrm{ and }
    S = 
    \left.    
    \frac{
      \partial \mathcal{M}(\vec{p})
    }{
      \partial \vec{p}
    } 
    \right|_{\vec{p}=\vec{p}_0} \,.
\end{equation}

Now we can compute the evidence, i.e., the probability distribution associated with $\vec{\sigma}_\textrm{exp}$.
This variable is multivariate normal because it is given as a sum of multivariate normal random variables.
Therefore it suffices to calculate the associated mean vector and  covariance matrix,
\begin{equation}
\label{eq:marpars}
\begin{gathered}
   \mathds{E}[\vec{\sigma}_\textrm{exp}] = \vec{\sigma}_\textrm{mod} \,,\\
  \textrm{Var}[\vec{\sigma}_\textrm{exp}] = M =
  S K_\textrm{par} S^T + K_\textrm{def} + K_\textrm{exp} \,,
\end{gathered}
\end{equation}
to obtain the result that $\vec{\sigma}_\textrm{exp} \sim \mathcal{N}(\vec{\sigma}_\textrm{mod}, M)$ which corresponds to the probability density function~$\pi(\vec{\sigma}_\textrm{exp})$ for a specific realization $\vec{\sigma}_\textrm{exp}'$ given in~\cref{eq:logmarlike}.
Noteworthy, this distribution provides prior probabilities of possible outcomes of the experiment.
As an aside, we remark that the underlying statistical computation---called marginalization---has been used for nuclear data evaluation before~\cite{cyrilledesaintjean_monte_2009}.
Marginalization serves the purpose to get rid of variables not of interest by themselves but nevertheless to properly account for the extra uncertainty they introduce.
Here we marginalized over nuclear model parameters.

As soon as we have made the experimental measurements, we know the values in $\vec{\sigma}_\textrm{exp}$.
This vector of experimental measurements must be compatible with the theoretical distribution parameters in \cref{eq:marpars}.
Without the covariance matrix $K_\textrm{def}$ it is usually not because too many things can go wrong.
For instance, some error sources may have been overlooked for the construction of the experimental covariance matrix or the model is not able to adapt perfectly to the experimental data.
Acknowledging the fact that we need statistical compatibility between the realization of $\vec{\sigma}_\textrm{exp}$ coming from real experiments and the a priori theoretical distribution, we remove the inconsistency by adjusting the structure of $K_\textrm{def}$. 
We achieve this by adjusting the parameters $\{u_i\}_{i=0..N}$ and $\{v_i\}_{i=1..N}$ of \cref{eq:transAmpAndMetric} which define the form of \cref{eq:dtwcovfun} and consequently also $K_\textrm{def}$ to maximize $\pi(\vec{\sigma}_\textrm{exp}')$,
\begin{align}
\label{eq:logmarlike}
  \log \pi(\vec{\sigma}_\textrm{exp}') =
  -\frac{N}{2} \log (2\pi) 
  -\frac{1}{2} \log |M|
  -\frac{1}{2} \chi^2 \,, \textrm{ with } \\
  \chi^2 = \left(
    \vec{\sigma}_\textrm{exp}' -
    S \vec{\sigma}_\textrm{mod}  \right)^T
  M^{-1}
    \left(
    \vec{\sigma}_\textrm{exp}' -
    S \vec{\sigma}_\textrm{mod}    
  \right) \,.
\end{align}
In other words, we seek to adjust $\delta(.)$ and $m(.)$ of the covariance function  to maximize the probability of the experimental measurements~$\vec{\sigma}_\textrm{exp}'$.
Noteworthy, a linear transformation of the occurring vectors and corresponding transformations of $S$ and $M$ leave the value of the third term (i.e. the $\chi^2$ term) invariant but alter the value of the second term containing the determinant.
It seems more reasonable to us to penalize complexity on a relative scale and hence we used the transformation $\tilde{\sigma}_i = (\sigma_i - \sigma_{\textrm{mod},i}) / \max(0.1, \sigma_{\textrm{mod},i})$ to transform $\vec{\sigma}_\textrm{exp}$ and $\vec{\sigma}_\textrm{mod}$ and also performed the corrsponding transformation on the sensitivity matrix $S$ and the covariance matrix $M$.
We expect the model to be less predictive for very low cross sections.
To avoid unreasonable large uncertainty bands for such cross sections, we revert back to an absolute scale for cross sections below 0.1 millibarn, which explains the maximum in the denominator.
The model defect is applied additively on this transformed scale.
The amplitude function $\delta(E)$ therefore roughly amounts to the relative deviation of the truth from the model if the underlying prior model cross section is greater than 0.1 mbarn. 

We are interested to learn about the \textit{global} predictive power of the nuclear physics model.
Thus we take reaction data of several neighboring isotopes into account for the maximization of \cref{eq:logmarlike}.
We aim to find a set of $\{u_i\}_{i=0..N}$ and $\{v_i\}_{i=0..N}$ that is suited for a specific reaction channel, e.g., (n,p), for all the isotopes.
Assuming experimental data and model parameters associated with different isotopes to be independent, we obtain the following form of the marginal likelihood $\pi(\vec{\sigma}_\textrm{exp}')$:
\begin{equation}
\label{eq:sumlogmarlike}
  \log \pi(\vec{\sigma}_\textrm{exp}) =
  -\frac{N}{2} \log(2\pi)
  -\frac{1}{2} \sum_{i=1}^{N_\textrm{iso}} \log |M_i| 
  +\log\left( \sum_{i=1}^{N_\textrm{iso}} \exp\left(
    -\frac{1}{2} \chi_i^2 
  \right)\right)
\end{equation}
The sums run over the number of isotopes $N_\textrm{iso}$ and contain the covariance matrices $M_i$ and $\chi^2$ values with respect to the individual isotopes.

\section{Exemplary application}
\label{sec:example}
\begin{figure*}
\includegraphics[width=\textwidth]{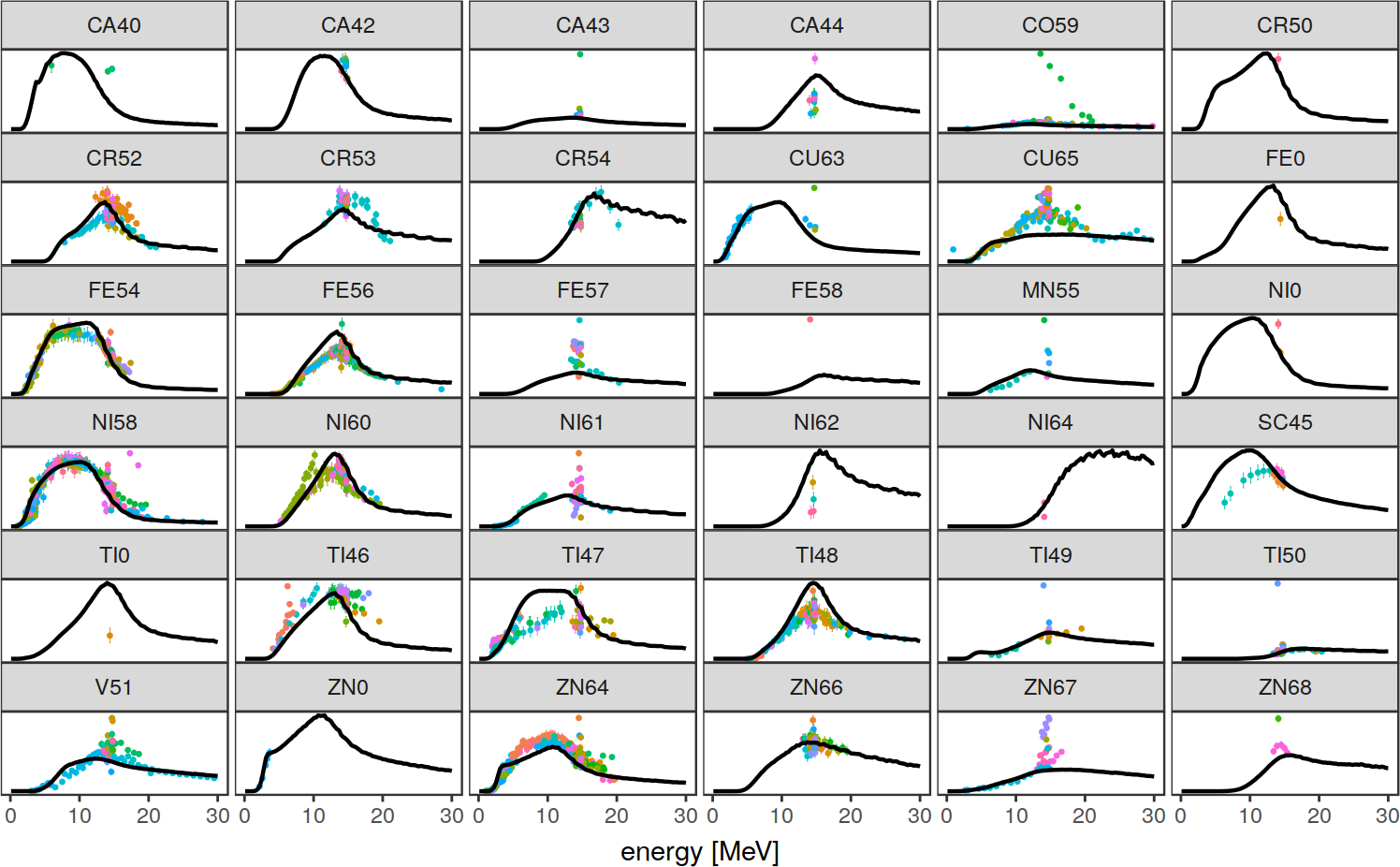}
\caption{Experimental data for angle-integrated (n,p) cross sections of isotopes with charge numbers between 20 and 30 used for the marginal likelihood maximization.
Uncertainties of experiments are assumed to be $10\%$ and independent across data points.
An energy of zero corresponds to the threshold energy of the reaction.
The mass number zero indicates natural composition.
No efforts have been made to identify outliers or assess the quality of the data sets.
The data were retrieved from 207 entries of the EXFOR database.}
\label{fig:modefSys}

  \begin{minipage}{0.32\textwidth}
    \centering
    \includegraphics{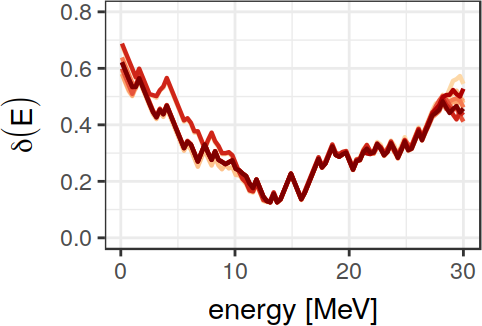}
  \end{minipage}
  \hfill
  \begin{minipage}{0.32\textwidth}
    \centering
    \includegraphics{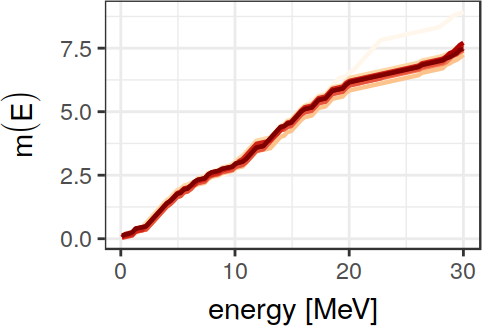} 
  \end{minipage}
  \hfill
  \begin{minipage}{0.32\textwidth}
    \centering
    \includegraphics{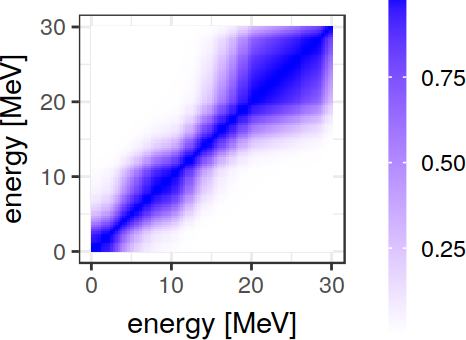} 
  \end{minipage}
  \caption{The dynamic time warping GP specification resulting from the marginal
   likelihood maximization:
  (left) the amplitude function $\delta(.)$; (middle) the metric function $m(.)$; and (right) the corresponding model defect correlation matrix.
Different curves in the left and middle figure correspond to different local maxima of the marginal likelihood.
The majority of the ten optimization results coincide and indicate the global maximum.
Due to the transformation of the cross sections described in the paragraph below~\cref{eq:logmarlike}, the function $\delta(E)$ can be interpreted as relative model defect expected a priori.
}
\label{fig:optResults}

\end{figure*}

We determined the model defect covariance function by the outlined approach for (n,p) reactions in combination with the nuclear models code TALYS~\cite{koningTALYS1NuclearReaction2013}.
Because we want to use the model defects for an evaluation of $^{56}$Fe, we included angle-integrated cross section data for isotopes with charge numbers between 20 and 30 and incident energies between 0.1 and 30\,MeV available in the EXFOR database~\cite{otuka_towards_2014}.
We assigned an uncorrelated uncertainty of 10\% to all the data points, which is sufficient for the purpose of demonstrating the outlined approach. 
The included experimental data for (n,p) reactions is displayed in~\cref{fig:modefSys}.
As can be seen in the figure, we did not attempt to remove experimental outliers.
Unfortunately, given space restrictions, we are not able to give proper credit to the authors of these measurements due the fact that the data is spread over 207 entries in the database.
References to the data can be provided upon request.

We employed an equidistantly spaced energy grid with 100 points covering the range from 0.1 to 30\,MeV for the construction of $\delta(E)$ and $m(E)$ given in \cref{eq:transAmpAndMetric}.
This point density corresponds roughly to an energy resolution of 0.3\,MeV and gives rise to 200 hyperparameters (i.e. the $u_i$'s and $v_i$'s in~\cref{eq:transAmpAndMetric}) that have to be optimized.
Importantly, we shifted the origin of the energy grid to the threshold of the reactions.
This measure ensures that cross sections near thresholds of different isotopes are mapped to the same grid points of the Gaussian process and also peaks are potentially better aligned.

We used the L-BFGS-B algorithm~\cite{byrd_limited_1995} for the maximization of \cref{eq:sumlogmarlike}, which also permits the specification of box constraints.
Applying box constraints on the transformed variables in \cref{eq:transAmpAndMetric} enabled us to regularize the shape of the solution.
We enforced that the amplitude function may not change more than about three percent between consecutive grid points.
In other words, we constrained the rate of change of the amplitude function to be below roughly 10\% per MeV.
As the metric function has to be monotonically increasing, we applied a lower limit of 0.03 and an upper limit of 0.15 for its change between consecutive grid points.
This specification translates to the prior knowledge that the effective length-scale should be at least roughly 1\,MeV and not larger than 10\,MeV at any energy.
The lower limit on the effective length-scale protects against discrepant data, which would elsewise drive the effective length-scale to unreasonable low values.
We emphasize here again that the dynamic amplitude and length-scale determined in a data-driven way is the real novelty of our approach.
We derived analytic expressions for the gradient of the marginal likelihood, see, e.g.,~\cite{2018arXiv180300960S}, which can be exploited by the L-BFGS-B algorithm.
We performed the optimization ten times with different starting values and achieved in all cases convergence.
Averaged over the individual optimization runs, it took the algorithm about 630 iterations to converge---tens of minutes on a decent personal computer with eight cores.
Even though there were two local maxima, the majority of calculations converged to the global one.
The learned amplitude and metric function as well as the correlation matrix of the model defect are shown in~\cref{fig:optResults}.

\section{Summary and conclusion}
\label{sec:summary}
We introduced a new parametrization of the covariance function based on the concept of an amplitude and metric function and showed how these functions can be inferred from experimental data for neighboring isotopes by maximizing the marginal likelihood.
We demonstrated the feasibility of the approach with the nuclear models code TALYS~\cite{koningTALYS1NuclearReaction2013} and (n,p) reaction data for isotopes with charge numbers between 20 and 30.
The obtained amplitude function and metric functions are interesting by themselves because they potentially allow us to gain not only qualitative but also quantitative insight into model performance on a per-energy basis.
In the future, these model defects constructed by taking a global view on the predictive performance of nuclear models can be included in evaluation procedures to obtain more robust and reliable results.
  
%
% BibTeX or Biber users please use (the style is already called in the class, ensure that the "woc.bst" style is in your local directory)
\bibliography{WONDER18}

\begin{thebibliography}{14}

\bibitem{helgessonAssessmentNovelTechniques2017}
P.~Helgesson, D.~Neudecker, H.~Sjostrand, M.~Grosskopf, D.L. Smith, R.~Capote,
  \emph{Assessment of {{Novel Techniques}} for {{Nuclear Data Evaluation}}}, in
  \emph{{{ASTM}} 16th {{International Symposium}} on {{Reactor Dosimetry}}}
  (Santa Fe, New Mexico, 2017)

\bibitem{pigni_uncertainty_2003}
M.T. Pigni, H.~Leeb, \emph{Uncertainty {{Estimates}} of {{Evaluated}}
  {\textsuperscript{56}}{{Fe Cross Sections Based}} on {{Extensive Modelling}}
  at {{Energies Beyond}} 20 {{MeV}}}, in \emph{Proc. {{Int}}. {{Workshop}} on
  {{Nuclear Data}} for the {{Transmutation}} of {{Nuclear Waste}}.
  {{GSI}}-{{Darmstadt}}, {{Germany}}} (2003)

\bibitem{leeb_consistent_2008}
H.~Leeb, D.~Neudecker, T.~Srdinko, Nuclear Data Sheets \textbf{109}, 2762
  (2008)

\bibitem{schnabel_large_2015}
G.~Schnabel, Ph.D. thesis, Technische Universit\"at Wien, Vienna (2015)

\bibitem{rasmussen_gaussian_2006}
C.E. Rasmussen, C.K.I. Williams, \emph{Gaussian {{Processes}} for {{Machine
  Learning}}} ({MIT Press}, Cambridge, Mass., 2006), ISBN 0-262-18253-X
  978-0-262-18253-9

\bibitem{schnabel_differential_2016}
G.~Schnabel, H.~Leeb, EPJ Web of Conferences \textbf{111}, 09001 (2016)

\bibitem{2018arXiv180300928S}
G.~Schnabel, arXiv eprint arXiv:1803.00928  (2018)

\bibitem{helgessonFittingDefectNonlinear2017}
P.~Helgesson, H.~Sj\"ostrand, Review of Scientific Instruments \textbf{88},
  115114 (2017)

\bibitem{myersPerformanceTradeoffsDynamic1980}
C.~Myers, L.~Rabiner, A.~Rosenberg, IEEE Transactions on Acoustics, Speech, and
  Signal Processing \textbf{28}, 623 (1980)

\bibitem{cyrilledesaintjean_monte_2009}
{Cyrille De Saint Jean}, G.~Noguere, B.~Habert, B.~Iooss, Nuclear Science \&
  Engineering \textbf{161}, 363 (2009)

\bibitem{koningTALYS1NuclearReaction2013}
A.J. Koning, S.~Hilaire, S.~Goriely, \emph{{{TALYS}}-1.6 - {{A Nuclear Reaction
  Program}}}, http://www.talys.eu (2013)

\bibitem{otuka_towards_2014}
N.~Otuka, E.~Dupont, V.~Semkova, B.~Pritychenko, A.~Blokhin, M.~Aikawa,
  S.~Babykina, M.~Bossant, G.~Chen, S.~Dunaeva et~al., Nuclear Data Sheets
  \textbf{120}, 272 (2014)

\bibitem{byrd_limited_1995}
R.H. Byrd, P.~Lu, J.~Nocedal, C.~Zhu, SIAM Journal on Scientific Computing
  \textbf{16}, 1190 (1995)

\bibitem{2018arXiv180300960S}
G.~Schnabel, arXiv eprint arXiv:1803.00960  (2018)

\end{thebibliography}

\end{document}